# Active learning in pre-service science teacher education

Vera Montalbano, Roberto Benedetti

Department of Physical Sciences, Earth and Environment, University of Siena, Siena, Italy

Abstract – date of last change of the paper: 03.05.2014

*We report a course on teaching in physics lab for teachers enrolled in Formative Active Training, which actually allows to obtain the teacher qualification in Italy. The course was designed with the purpose of showing in practice what means active learning in physics and how effective activities can be realized. Two different type of teachers attended to the course, a small group, with physics or mathematics degree, for teacher qualification in secondary school of second grade (age 14-19) and a more numerous group for qualification in secondary school of first grade (age 11-14), usually with a different science degree such as biology, environmental sciences and so on. We compare the training in physics lab between the two groups and with other experiences we performed in previous years in pre-service education and updating courses for teachers in-service.*

Keywords: Lifelong learning, active learning in physics laboratory, Secondary education: lower (ages 11-15)

**Introduction**

Physics plays a fundamental role in science education as an accessible context for experimental design, scientific argumentation, problem solving, and the development of multi-step reasoning skills. Especially in the physics lab, students can actively develop scientific processes and mind habits typical of physics and science in general. However, undergraduate and graduate students in higher education have limited opportunities to experience topics meaningful for secondary education. Therefore, an unavoidable aim of pre-service education is to improve and develop teachers' skills in this direction. The most effective way seems to propose carefully design sequences of active physics learning in laboratory [1-4]. Thus, teachers can have a direct experience of the powerful support in comprehension of physical concepts and laws that can derive from active learning. At the same time, they can test in the laboratory some relevant experiments for teaching in secondary school.

After a period of pre-service education lack, the first course in Formative Active Training for obtaining teacher qualification started in Italy one years ago. In the following, we present the context in which an innovative course on Physics Laboratory Didactics was designed and realized. The participants were characterized by mean of an initial questionnaire, as shown in the next section. Actions for promoting active learning in science in designing and realizing the course and methods for assessment are given in the successive section. Finally, some preliminary results are presented and discussed.

Many laboratory activities were inspired by laboratories realized in an effective way [4] within the Italian National Plan for Science Degrees [5, 6].

**National Plan for Science Degrees and pre-service science teachers education**

In the last years, the decline of students' interest in learning physics and the consequent decrease of enrolments in physics in Italy have been contrasted by the Ministry of Education and Scientific Research through the promotion of a wide national plan [5] (Piano nazionale per le Lauree Scientifiche, i.e. PLS).

The main PLS actions have been professional development for teachers and students orientation, essentially through laboratory activities focused on orientation to a science degree by training and considering laboratory as a method not as a place. Student is considered the main actor of learning and joint planning by teachers and universities is encouraged.

**An active learning path on active learning**

Despite a long experience in pre-service education where focus was on improving disciplinary contents and teachers' competences, we fully realized such a powerful tool active learning can be only when we were engaged in realizing effective laboratories within PLS. In this context, we utilized a summer school of physics for a pre-service training and for professional development of young teachers [7, 8].

A selection of more effective activities developed in PLS [4,7,8] was the starting point for designing a learning path for teachers enrolled in Formative Active Training course. The aim was to engage teachers directly in active learning on meaningful topics, such as introducing to measure and evaluation of uncertainties. They worked in small groups, often in an inquiry-based activity performed sometimes in conditions very similar to those usually found in schools (few and poor materials, missing or ill-equipped laboratories). The next step was to render teachers aware of which activities had been effective and the active role played by the teacher (one of the authors) in favoring this achievement, i.e. a metacognitive reflection on the activity was encouraged.

Preliminary analysis on assessment shows that the goal of the learning path seemed achieved for science teachers in secondary school of first grade. On the contrary, mathematics and physics teachers in secondary school of second grade were still too focused on disciplinary contents and less aware of active learning.

**Pre-service science teacher education context**

In order to describe how pre-service science teacher education is evolved to the actual organization, let us give a brief survey of recent reforms on this issue in the last years.

**A brief history of teaching qualification in Italy**

For decades, there was no pre-service teacher education in Italy. All university graduates in a disciplinary degree could participate to a professorships by a competitive examination and obtain a teaching qualification and a permanent position at school.

In 1999, the Advanced School for Teaching in Secondary Schools (Scuola di Specializzazione all'Insegnamento Secondario, SSIS) became the only way for obtaining qualification for teaching in secondary schools of first and second grade. A limited number of students were admitted by exam and training at school was introduced. The SSIS management was regional and teaching sites were distributed in each university (in Tuscany at Pisa, Florence and Siena).

From the beginning, a team of mathematicians, physicists and expert teachers (both authors too) were involved in all teaching sites and elaborated together an effective educational program for SSIS of Tuscany in pre-service mathematics and physics teacher in secondary school of second grade. On the contrary, mathematician and physicists involvement in science teacher education was marginal because of prevalence of life sciences researchers.

After ten years of activity, SSIS was closed awaiting a new pre-service education course, i.e. Formative Active Training (Tirocinio Formativo Attivo, TFA) that finally started the last year.

Table 1. Teacher education in Italy in last decades

|  | Admission degree | Adv Course | Adm exam | Pre-service training | Teaching Qualification |
|---|---|---|---|---|---|
| Before 1999 | disciplinary degree | none | no | none | Professorships competitive examination for qualified participants |
| From 1999 to 2009 | disciplinary degree | biannual SSIS | yes | 290 hours exp teach | Exam for teaching qualification written and oral exam |
| 2012 | disciplinary degree | annual TFA | yes | 475 hours exp teach | Exam for teaching qualification final report on training and oral exam |
| Next future | teach. disc. degree | annual TFA | yes | 475 hours exp teach | Exam for teaching qualification final report on training and oral exam |

The main steps in reforming pre-service teacher education are summarised in table 1. In the next future teaching disciplinary degree will precede TFA.

**Formative Active Training framework**

The main educational program in TFA is outlined in table 2, where students´ work is assessed by credits, not present in SSIS. Both curricula for science teachers and mathematics and physics teachers are presented. The actual credit system allows less time to each course. For example, in SSIS Math & Phys teachers have activity for 4 full afternoons per week (1 for pedagogical course, 1 for mathematics and 2 for physics if they had a math degree) for four semesters. In TFA, for the same teaching qualification a student is occupied for 3 afternoons per week for slightly more than one semester.

Table 2. Formative Active Training education program

| Pedagogical competences | 18 credits | Education Science  12 credits  Education Science for special needs  6 credits | |
|---|---|---|---|
| Disciplinary contents | 18 credits | Math & Phys  Math Didactics  6 credits  Math Education  3 credits  Phys Lab Dida  3 credits  Classic & Mod Phys lab  6 credits | Math & Sciences  Math Didactics  6 credits  Phys Lab Dida  3 credits  Chem. Lab  3 credits  Bio Lab  3 credits  Earth Sc Lab  3 credits |
| Observative and active training at school | 19 credits | training under supervision of an expert teacher at school  400 hours teaching practice education  75 hours dedicated to students with special needs | |

The real novelty in TFA is the strong reinforcement of training at school, underlined by the relevance of a final report on training (see table 1) that must be presented and discussed by candidates in the examination for teaching qualification. Moreover supervision on final report is requested by an expert teacher and by an university supervisor.

All public universities of Tuscany decided to affiliate in order to maintain the regional coordination. The small number of students allowed by the Ministry implied to reduce teaching locations. Thus, for Math & Science teaching in secondary school of first grade the TFA management is at the University of Siena with teaching locations in Siena and Pisa. For Math & Phys teaching in secondary school of second grade the management is at the University of Pisa with teaching locations in Pisa, Florence and Siena.

Admission examination for TFA was done in 2012, courses started in 2013 (on February) and examination for qualifications was held in July. In Siena, TFA disciplinary courses for Math & Science were borrowed by other TFA courses.

**On characterizing in-training teachers**

The learning path on active learning was attended by Math & Science teachers (34) and Math & Phys teachers (11) in Siena. Since it was easy to collect data (final reports and examinations) for Math & Science teachers (26) from Pisa, we can consider them like a control group because no active learning was introduced in the physics course in Pisa.

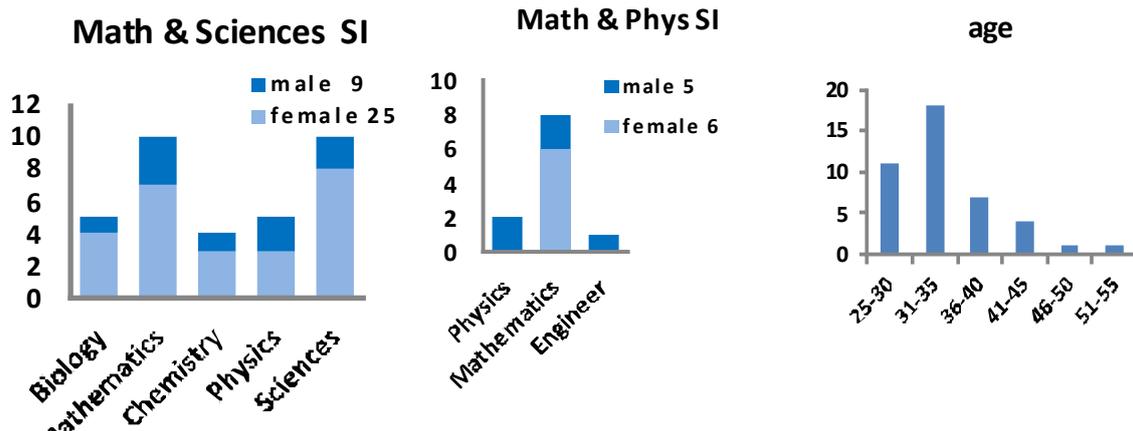

Figure 1. Types of disciplinary degree owned by participants separated by sex and teaching matter on the left and centre, participants age distribution on the right.

In fig. 1, the distributions of disciplinary degree of participants and their age are shown.

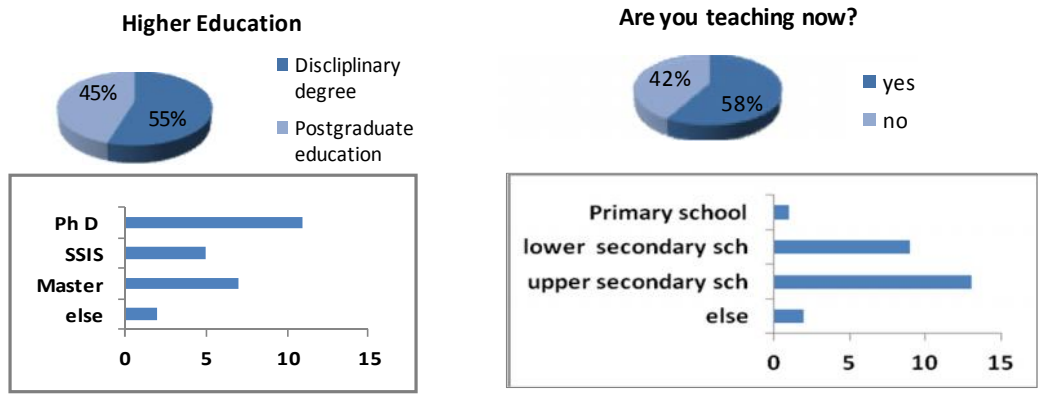

Figure 2. Participants' higher education on the left and actual position at school on the right.

The number of women is about double compared to men. The most of them have a degree in mathematics. There is a numerous group of people which have obtained their degree recently, but many got it five, ten and even fifteen years ago.

In the first lesson a questionnaire was completed by all participants in Siena, in order to have more information on the previous education and on their teaching experience. In fig. 2, previous higher education of participants is shown. Moreover, the most part (58%) were working at school with a temporary position.

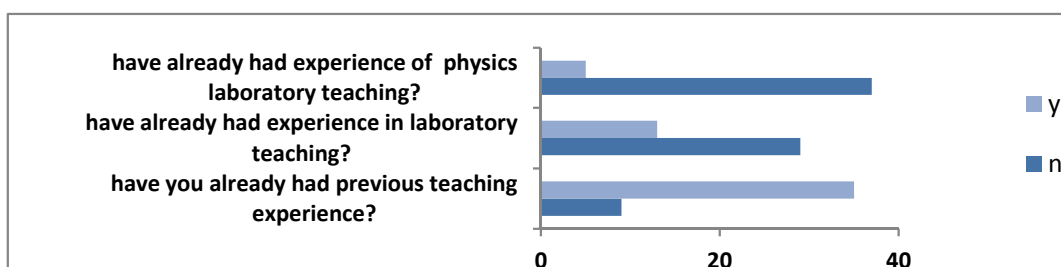

Figure 3. Participants' experience in teaching.

A set of questions is dedicated to previous experiences in teaching. As shown in fig. 3, although 83% just have had at least one teaching experience, only 12% have got one in physics laboratory.

## Actions for promoting active learning in science teacher education

The active learning path was carried out in a 30 hours course entitled "Physics Lab Didactics".

### Designing and realizing a learning path on active learning

Since the course was attended by two different kinds of students, it was necessary to split the class into two groups for many activities in laboratory in order to give for secondary schools of second grade some example of disciplinary lab useful for the last years of high school. The complete articulation of the course and disciplinary contents are given in table 3.

Table 3. Physics Lab Didactics: organization and contents

|  | Math & Science | Math & Phys |
|---|---|---|
| lessons with discussion 8 h | how to work in lab, safety management, how to organize a lab, how to write an effective lab report, a survey on reformed school, role of interdisciplinary in math/sciences, introduction to measure, etc. | |
| laboratory 22 h | 1. introduction to measure (mass, volume, density, direct and indirect measurements)<br>2. Measures of times (pendulum)<br>3. A qualitative path on friction | |
|  | 4a Introduction to sky observation<br>5a Orders of magnitude, estimates, measures<br>6a Qualitative and quantitative lab | 4b A quantitative path on friction<br>5b Measurements through video and Picture analysis<br>6b Calorimetric meas.<br>7b Electron's electric charge meas.<br>8b Measures of lengths by using light |

In laboratory, participants worked in small groups (4 components) with minimal initial instruction in order to be introduced to physics lab and measure, evaluation of uncertainties in measurements, measures of some basic physical quantities (act 1, 2, 3). A topic was focused on an experimental situation which can be suitable for students of different ages, and in this case the two groups faced different experiments (e. g. qualitative exploration on friction [9] was proposed to both groups but the quantitative lab [9] was performed only for secondary school of second grade). A great attention was put in rendering teachers aware of which activity could be effective in improving active learning and reflecting about the role assumed by the teacher in getting this achievement (meta cognitive elaboration).

Special care was put in introducing active learning examples, focus on behavior that can facilitate or inhibit it in lessons and especially in lab designing and execution. Many participants had few or no experience at all in phys lab or so few topics were discussed in university course in their degree so that active learning in disciplinary contents was really effective. Also physicists had usually such a deep specialized and different background that could discover a lot of unexpected details in direct experience in lab in the proposed basic topics. Groups were formed in an inhomogeneous way in order to stimulate cooperative learning [10]. Focus was put on their engagement, how to work in a group, how cooperative learning can be checked and facilitated.

As it is usual in the laboratory, everything could go in a wrong or unexpected way. These were the cases in which it was useful to underline how discussions can arise in groups and how to interact with students (a lot of good examples usually happens in lab and we discussed together how to manage them).

**Assessment**

For the assessment of the course, participants presented before the exam two reports on lab activity (an explicit request was that reports were written for peer readers) with a brief final educational discussion or one report and a proposal for an active learning activity in lab for a well-defined class. Reports where discussed in an informal way, before exam, focusing on active learning aspects.

Table 4. Methods to assess the effectiveness on participants

| Methods & Materials | Participants Math & Sc SI | Participants Math & Sc PI | Participants Math & Phys SI | state |
|---|---|---|---|---|
| 2 Lab reports with did. analysis or 1 lab report + 1 proposal focused on active learning in phys lab | 34 | - | 11 | done |
| Interviews on lab reports | 34 | - | 11 | done |
| Final report on training | 34 | 26 | - | in progress |
| Oral examination for teaching qualification | 32 mixed PI/SI | | - | in progress |

In table 4, all possible assessments are presented. Final reports acquisition and oral examinations in final exam will be completed for few last students at the end of October.

**Preliminary Results**

The first two methods of assessment showed in table 4 have been completed and the results are summarised in fig. 4. The learning path seemed more effective for Math & Science teachers respect to Math & Phys teachers.

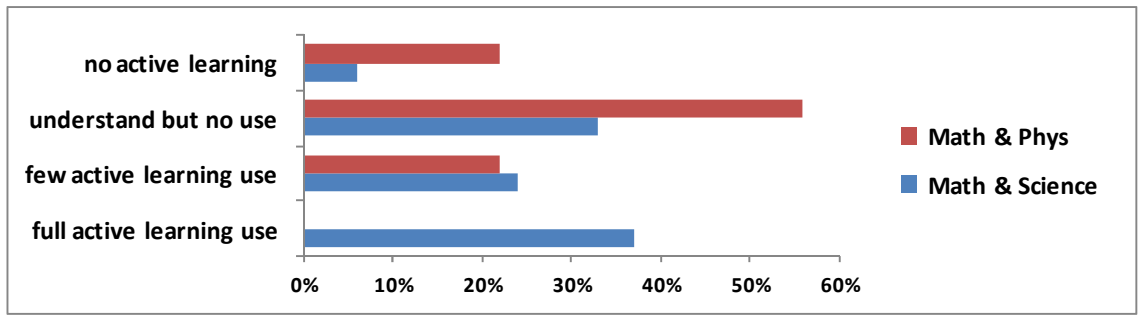

Figure 4. Results from lab reports and interviews for the two groups of participants.

Math & Phys teachers remained still too focused on disciplinary details and some of them seemed to have understood what active learning is only during the discussion in the exam of the course.

**An example of teacher's elaboration**

Many secondary school teachers of first grade showed a personal elaboration on active learning in proposals of other learning paths, in their final reports on training and in the oral exam for qualification. Some teachers tested successfully new learning paths in their training at school. Others use properly active learning in elaborating learning paths in sciences different from physics.

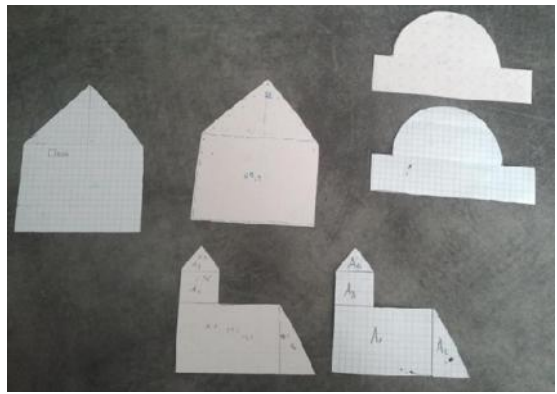

Figure 5. Examples of areas proposed in class for introducing the concept of measure, evaluation of uncertainties and measurements with different sensibility. Figures can be recognized like composed by adding simple geometrical figures, like triangles, squares and so on. Direct and indirect measures of areas can be done and compared.

A teacher presented her experience in training at school in which an introduction to measure was realized by proposing measurements of an area by direct comparing with different units of area (by using paper with a size grid of mm or cm). Not integer units must be estimated in different ways by students starting an interesting discussion which bring to introduce in a correct way the concept of uncertainty. Another way can be to measure lengths and perform a calculus (how can uncertainty be estimate in this case?). The subject was proposed in the course, but some actions, such as to propose area for measurements in the form of stylized buildings for a better motivation of students (see fig. 5) or to explain how different units can be relevant to achieve a more precise measurement by measuring the blackboard by means of sheets of different sizes, was proposed by the teacher.

**Conclusions**

Sharing expertise and creating knowledge in a group is a continuous process, in which members must be aware of their roles and how to monitor the work in an effective way. Some experiences of active learning in physics laboratory followed by metacognitive reflection on the role of teacher in favouring this process seem to be useful in pre-service teacher's education. Moreover, some secondary school teachers of first grade transferred active learning directly in training at school and in teaching other sciences, some students became enthusiastic for active learning, others began to enjoy physics and phys lab.

From preliminary results it is possible to outline that physics lab designed for promoting active learning can be useful in inducing a deeper awareness on this issue. Even though, few teachers failed to distinguish between activity in the laboratory and active learning. Secondary school teachers of second grade remained still too focused on lab skills and disciplinary details and a careful reflection must be done in order to propose a different organisation of TFA courses and more effective activities in this case.